# Atmospheric Modelling for Neptune's Methane D/H Ratio – Preliminary Results

Daniel V. Cotton, Lucyna L. Kedziora-Chudczer, Kimberly Bott and Jeremy A. Bailey

*School of Physics, University of New South Wales, NSW 2052, Australia*

**Summary:** The ratio of deuterium to hydrogen (D/H ratio) of Solar System bodies is an important clue to their formation histories. Here we fit a Neptunian atmospheric model to Gemini Near Infrared Spectrograph (GNIRS) high spectral resolution observations and determine the D/H ratio in methane absorption in the infrared H-band (~1.6 μm). The model was derived using our radiative transfer software VSTAR (Versatile Software for the Transfer of Atmospheric Radiation) and atmospheric fitting software ATMOF (ATMOspheric Fitting). The methane line list used for this work has only become available in the last few years, enabling a refinement of earlier estimates. We identify a bright region on the planetary disc and find it to correspond to an optically thick lower cloud. Our preliminary determination of $CH_3D/CH_4$ is $3.0 \times 10^{-4}$, which is in line with the recent determination of Irwin et al. [1] of $3.0 \times 10^{-4}$ ($^{+1.0}/_{-0.9} \times 10^{-4}$), made using the same model parameters and line list but different observational data. This supports evidence that the proto-solar ice D/H ratio of Neptune is much less than that of the comets, and suggests Neptune formed inside its present orbit.

## Introduction

**Deuterium as an indicator of planet formation location**

The amount of deuterium in the universe is decreasing. Deuterium was produced in the Big Bang, and is destroyed in stars [2]. The amount of deuterium present during the formation of the Solar System is referred to as the *proto-solar D/H ratio*. The atmospheres of Jupiter and Saturn are considered representative of the proto-solar D/H ratio [1]; they are thought to have formed first of the giant planets at the *water ice-line*, representing the distance from the Sun at which water freezes – from the accretion, predominantly of water ice and clathrate hydrates [1]. Although much of the water ice in the proto-solar nebula is derived from the dense interstellar medium, its initial D/H ratio is not wholly retained; reprocessing takes place as part of the formation of the solar nebula, altering the D/H ratio [3]. Beyond the water ice-line the temperature controls the D/H ratio in water ice through an equilibrium reaction [2]:

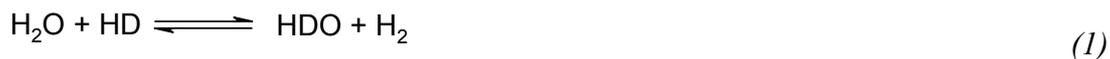
$$H_2O + HD \rightleftharpoons HDO + H_2 \qquad (1)$$

Further away from the Sun the colder temperatures result in more deuterium being found in heavy-water ice rather than deuterated molecular hydrogen. The giant planets are thought to have accreted from the proto-solar nebula around icy cores formed predominantly from water ice [4]. Consequently, the D/H ratio of a planet – either in water, the accreted (predominantly Hydrogen) gas envelope or products evolved from those – is indicative of where it formed in relation to the Sun. Where the atmospheric constituents are evolved from the originally accreted material, measurement of these can be used to determine the D/H ratio of the accreted material that formed the planet. Determinations made of the D/H ratio of Jupiter and Saturn rely on an assumption – backed by models of the interiors of the planets – that deuterium present in the core ices have not significantly enriched the outer gas envelope [5]. In contrast, Uranus and Neptune are thought to have gas envelopes enriched through mixing during the formation by volatiles from their ice cores or infalling planetesimals; thus the comparison with Jupiter and Saturn provides an estimate of the deuteration of the ices in the local environment of their formation [5]. In the case of Neptune, methane is deuterated to its present level through an isotopic exchange reaction with molecular hydrogen [5]:

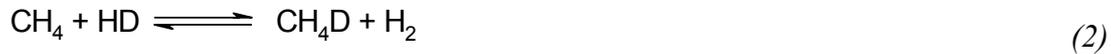

$$CH_4 + HD \rightleftharpoons CH_3D + H_2 \qquad (2)$$

Recently it has been proposed that Uranus and Neptune formed at the carbon monoxide ice-line, and thus were formed predominantly from carbon monoxide ice [6]. If this is the case, then the water in the interior of these planets will have CO as its origin, with a D/H ratio representative of that origin. CO in the core would react with $H_2$ according to [6]:

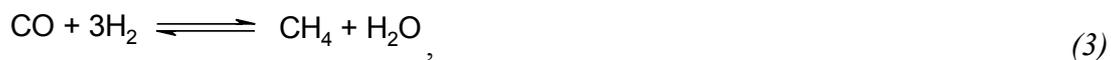

$$CO + 3H_2 \rightleftharpoons CH_4 + H_2O, \qquad (3)$$

and then as a greater portion of deuterium in the atmospheres of Uranus and Neptune would have come from the hydrogen envelope, the D/H ratio would be expected to be lower than for formation from water ices the same distance from the Sun. Recent measurements of the D/H ratios of Uranus and Neptune are lower than those of comets, which are thought to have formed in the same region of the Solar System, and so Ali-Dib et al. suggest that a CO ice origin reconciles this position [6]. Ali-Dib et al.'s initial calculations of the D/H ratio reveal a discrepancy with the measured ratio in that they infer a $H_2$/He ratio different to the Solar ratio during their formation [6]. A definitive measurement of the D/H ratio, for both Neptune and Uranus (which we are also investigating [7]) is very important to the viability of their hypothesis.

**Recent measurements of the Neptunian D/H ratio**

The most recent measurement of Neptune's D/H ratio was made by Irwin et al. [1]. They used the Near-infrared Integral Field Spectrometer (NIFS) on Gemini to obtain R = 5290 infrared H-band spectra to measure the $CH_3D/CH_4$ ratio from the $CH_3D$ $3\nu_2$ lines. Only recently have sufficiently detailed line lists become available to accurately model methane at cryogenic temperatures [8]. Previously Feuchtgruber et al. [9] used the Photodetector Array Camera and Spectrometer (PACS) instrument on Herschel to determine the D/H ratio in molecular hydrogen from its R(0) and R(1) HD lines, supplemented by previous measurements of the R(2) rotational line using the Short Wavelength Spectrometer on board the Infrared Space Observatory [4]. To compare measurements of the D/H ratio in $H_2$ to those of the $CH_3D/CH_4$ ratio requires a knowledge of the fractionation factor, which describes how the D/H ratio differs in chemical products of the original material. The fractionation factor has been estimated by Lecluse et al. [5], and used by Irwin et al. to make the comparison: the results are summarised in Table 1.

*Table 1: Recent Measurements of the Neptunian D/H ratio. Italicised entries represent calculations based on the fractionation factor estimated by Lecluse et al. (1.61+/-0.21) [5].*

| Reference | Instrument | D/H (H$_2$) (x10$^{-5}$) | | CH$_3$D/CH$_4$ (x10$^{-4}$) | |
|---|---|---|---|---|---|
| Feuchtgruber et al. 2013 [9]. | Herschel-PACS | 4.1 | +0.4 | 2.6 | *+0.4* |
|  |  |  | -0.4 |  | *-0.4* |
| Irwin et al. 2014 [1]. | Gemini/NIFS | *4.7* | *+1.7* | 3.0 | +1.0 |
|  |  |  | *-1.5* |  | -0.9 |

**Neptunian Clouds**

Neptune has a dynamic atmosphere with the highest zonal wind speeds in the Solar System (400 m/s) and a variety of cloud features [10]. In contrast to Neptune's uniform appearance in the visible, at infrared wavelengths there can be a striking contrast between light and dark features, where the lighter features correspond to haze in the stratosphere (~0.01 – 0.1 bar) and clouds near the tropopause (~2 – 3 bar) [10, 11]. What constitutes the haze and cloud layers seen in the (1.6 μm) H-band is not well understood [1] and although various trace hydrocarbons detected in the mid-infrared would condense out at pressures of around 0.002 – 0.01 bar their optical depths are calculated to be negligible in the H-band [10].

## Observations and Data Reduction

**Observations**

We made observations of Neptune on the 18th of August 2011 using the Gemini Near Infrared Spectrograph (GNIRS) instrument on the Gemini-North 8m telescope, using a 0.1 arc-second slit. H-band (1.46 – 1.63 μm) spectra were obtained using both the 32 and 110.5 lines/mm gratings, having a resolution of R = 5100 and R = 17800 respectively at 1.65 μm. The higher resolution data represents a narrower spectral window (1.525 – 1.57 μm).

An acquisition image of Neptune was taken before each long-slit observation, with the cloud pattern appearing nearly identical. As such, we have indicated in Fig. 1 the slit positions and extracted pixels of both observations on one of the acquisition images. The slit was orientated N-S on the sky, an acute angle to planetary north (as shown by reference to the north-pointing arrow in Fig. 1). The 1-D spectra are obtained from the average of 5 pixels along the 0.1 arcsec slit corresponding to the maximum signal area, which overlaps with the brightest patch of cloud in the acquisition image of the planet taken with the narrowband H-G0516 filter (shown in Fig. 1).

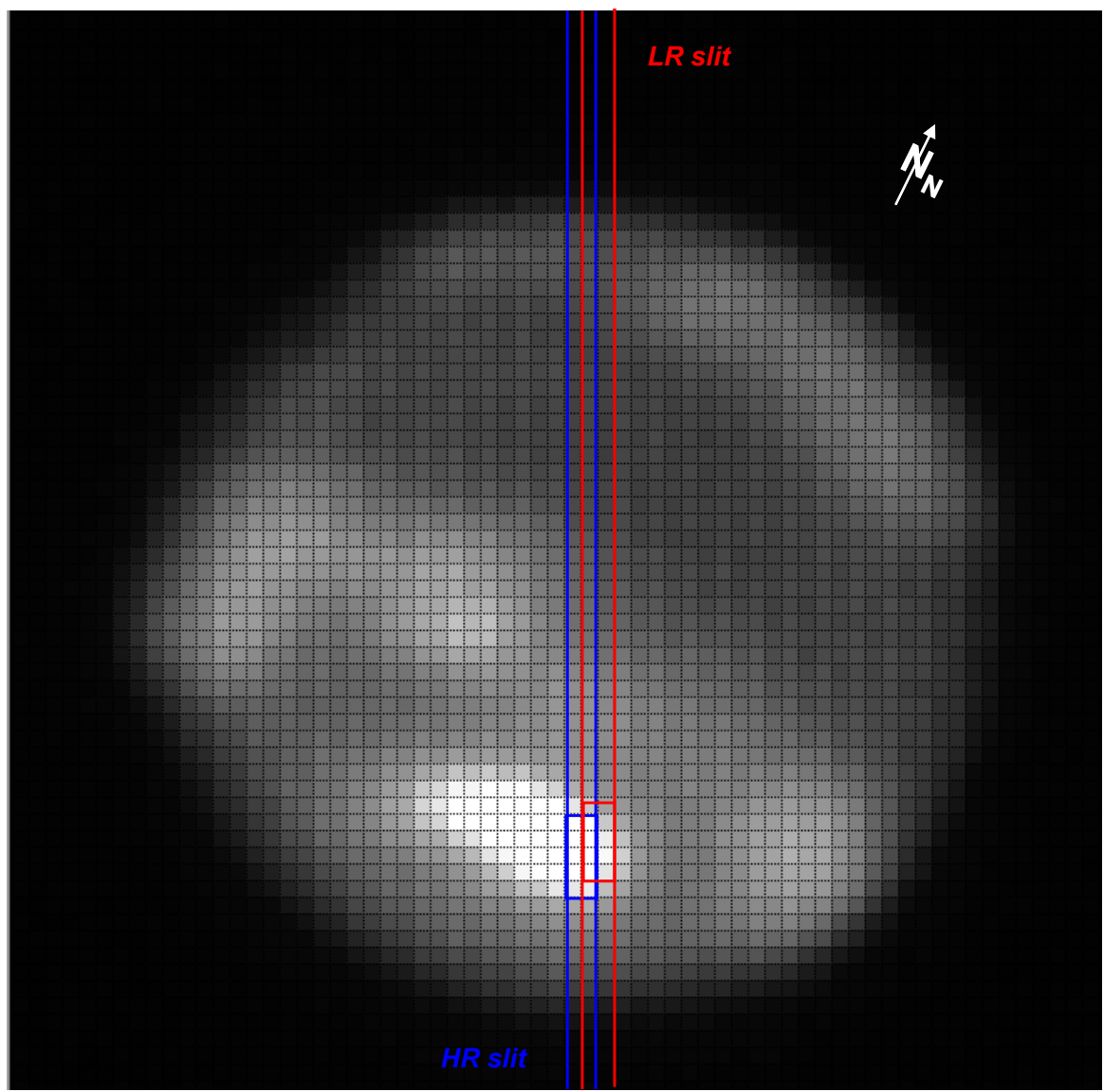

*Fig. 1 Acquisition image for the spectroscopic observations obtained; made with the narrowband H-G0516 filter. The positions of the slit and the extracted pixels are indicated for both the lower resolution (LR) and high resolution (HR) spectra. Image is scaled in greyscale, with white areas being the brightest, and black darkest. Planetary North is indicated in the top right.*

We fit an atmospheric model to both low-resolution* and high-resolution spectra. However in this work we present results of the fitting and retrieval of the lower resolution spectra only. The larger wavelength span of such a spectrum allows a robust determination of an appropriate cloud model that can be compared with previously published data. The best fitting model will then be applied to our high-resolution spectrum which will be reported on elsewhere.

At the time of the observation (07:23:27 UT) the mean airmass was 1.891, and the relative humidity 8%. For the low resolution spectra, our observing strategy involved twice obtaining four 350 sec exposures with Neptune moved between two slit positions (A and B) in an

---

* We refer to the R=5100 spectra as low-resolution only in comparison to the R=17800 spectra. R=5100 is higher resolution than was possible for this work prior to the availability of the current methane line list.

ABBA sequence for efficient sky subtraction, and averaging these. Wavelength calibration lamps (Ar-Xe) and spectra of the telluric standard star HIP 105734 (G1V type) were acquired after each sequence

**Standard Data Reduction**

The Gemini IRAF data reduction technique was applied to our spectra as follows: After visual examination of 2D spectra, all the data frames were aligned, corrected for bias level, non-linearity and bad pixels on the Aladdin III InSb detector array. The spectra of both Neptune and HIP 105734 were then flat fielded and wavelength calibrated using an interactive procedure to ensure correct identification of the Argon and Xenon lines present in the spectral region. The sky subtraction was performed by combining the A position frames, and subtracting the B position frames. The final step was to apply a dispersion correction and extract the 1D straightened spectra from the region of highest intensity along the 2D frame; this corresponded to the 5 pixel wide area (0.1x~0.25 arc-seconds) shown in Fig. 1.

**Telluric Removal**

The removal of the telluric absorption from our spectra was carried out using the ATMOF (ATMOspheric Fitting) code [12] as follows: We first take a high-resolution solar spectrum, and pass it through models of the Earth's atmosphere and the instrumental response. A number of model parameters are free, and ATMOF fits these to match the standard star data. The Earth atmosphere model used in this instance was the inbuilt VSTAR (Versatile Software for the Transfer of Atmospheric Radiation) [13] for Mauna Kea Observatory; its free parameters were the $CO_2$ and $H_2O$ content. $CO_2$ is varied as a multiplication of the whole altitudinal profile, whilst $H_2O$ is only varied for the lower layers of the atmosphere as detailed in Cotton et al. [12]. In the H-band water is a strong absorber at the short end of the band, and there is a smaller contribution from $CO_2$ near the centre of the band.

The instrumental response consisted of an augmented filter function, a filter response, a scaling factor and slope, as well as a wavelength shift correction. The augmented filter response was obtained from an average of the low frequency residuals from other standard star fits in the same wavelength window observed by us as part of the same observing run; the process is described in more detail by Bailey [14].

Once the parameters for the Earth atmosphere model have been retrieved, the model is re-run with the zenith angle corresponding to the Neptune observation. The Neptune spectrum is then multiplied by the Earth atmosphere transmission, and the instrumental response with the retrieved parameters applied. The only parameter retrieved from the standard star fitting not applied in correcting the Neptune model is the wavelength shift: this needs to be determined independently for each observation, as errors around 0.1Å can have a significant impact, and the quality of arc lamp calibration depends on the number of identified features in different spectral regions.

It is our preference to determine the wavelength shift by fitting a Doppler shifted solar spectrum to the planetary observational data; where the reflected solar lines are prominent in the spectrum this works well. However, this is not the case for Neptune in the H-band. Instead we used a basic model for Uranus from our earlier work [7] appropriately Doppler shifted. Neptune and Uranus have very similar spectra in the H-band, both dominated by methane absorption features, and as the purpose of this fitting is to retrieve only quadratic coefficients for wavelength shift, differences in the peak intensities or the overall flux are unimportant.

# Atmospheric Model and Fitting

To determine the D/H ratio we fit a Neptune VSTAR model to the spectrum with telluric lines removed using ATMOF.

**ATMOF Fitting**

Although we have previously demonstrated the ability of the ATMOF software and associated procedures to remove telluric features from our data [12], this is the first time it has been used as a retrieval scheme to derive model parameters for a planetary atmosphere. Here we begin with the processed (tellurics and instrumental response removed) planetary spectrum. We then modify parameters of an atmospheric model to produce a model spectrum and compare it with the data at the same spectral resolution (R) at which the data was collected. For this purpose we used the published values of R from the Gemini website[†].

It is customary to present planetary spectra as a *radiance factor*, i.e. the planet radiance divided by the incident stellar flux; we have not done this when fitting with ATMOF. Commonly the removal of telluric lines is approximated by division by a standard star spectrum. As we have more precisely removed the telluric lines using ATMOF, to present our spectra as radiance factor, we in fact have to divide by a solar spectrum. Any model we fit to spectra presented as radiance factor would also have to be divided by a solar spectrum (since it also includes light from the Sun as a direct beam source) which is a complication without benefit.

**Initial Neptune Model**

The parameters of the atmosphere model applied to our Neptune data primarily come directly from the most recent work of Irwin et al. [1]. Pressure and temperature profiles were taken from the results of the *Voyager 2* radio occultations through the planetary atmosphere [15]. The methane profile used is that by Irwin et al. [1]: the deep methane mole fraction is set to 4%, the maximum mole fraction limited to 60% saturation vapour pressure in the troposphere, but limited to $6 \times 10^{-4}$ in the stratosphere; a value used previously by Karkoschka and Tomasko [16], being intermediate of those derived by Lellouch et al. [17] and Baines et al. [18]. The $N_2$ mixing ratio was set to a constant 0.3% [1], and the remainder at each level divided amongst $H_2$ and He in the ratio 0.823/0.177 [15]. The atmospheric temperature, pressure, and gas mixing ratio profiles are shown in Fig. 2; they cover a pressure range between $3.5 \times 10^{-4}$ to 6.3 bar. The VSTAR, radiative-transfer calculations are performed by subdividing this range of pressures into 45 levels. VSTAR requires altitudes corresponding to pressure values, these were calculated according to,

$$\Delta z = \frac{T_z}{L_b}\left(\Delta P^{\frac{RL_b}{gM_z}} - 1\right) \quad (4)$$

where $\Delta z$ is the change in altitude; $T_z$ the temperature at that altitude; $\Delta P$ the change in pressure; $M_z$ is the molecular mass at height $z$, calculated using the mixing ratios of the gases in the model; $R$ the universal gas constant = 8.31432 N.m.mol$^{-1}$K$^{-1}$; $g$ the gravity of Neptune, taken to be 11.1046 m.s$^{-2}$[19]; and $L_b$ the adiabatic lapse rate, having a value of $8.53 \times 10^{-4}$ K.m$^{-1}$ [19]. The highest pressure in the model was then taken as zero altitude for reference.

---

[†] http://www.gemini.edu/sciops/instruments/gnirs/spectroscopy

The $CH_4$ line data used for the model are from laboratory measurements by the Grenoble group [20] made at cryogenic and room temperatures allowing the temperature dependence to be reliably determined. These line data have been shown to provide excellent models of the spectrum of Titan in the same wavelength region [2, 8, 21].

For the $CH_4$ lines broadened by $H_2$ we use the sub-Lorentzian line shape of Hartmann et al. [22], which has a far-wing $\chi$-factor of 0.05882. We include collision-induced absorption between both $H_2$-$H_2$ and $H_2$-He molecules, as well as Rayleigh scattering from $H_2$, He and $N_2$. Irwin et al. [1, 10] also included contributions from $H_2$-$CH_4$ and $CH_4$-$CH_4$ collision induced absorption, but the contributions from these should be less, and we have neglected them for the time being.

Our cloud model was based on Irwin et al.'s thin cloud model, which they found to provide a very similar fit to their more sophisticated continuous distribution of cloud [1]. The cloud model is a bi-layer cloud model based on their earlier work [10] with a lower cloud layer, and an upper cloud layer, referred to as a haze, with the pressure of the upper haze being consistent with pressures retrieved by Gibbard et al. for their clouds [11]. Both layers are calculated with a Henyey-Greenstein phase function with asymmetry factor 0.7 and a modified-Gamma log-normal size distribution of particles with an effective radius of 1 μm, $\sigma = 0.05$, and the refractive index of the particles is set to $1.4+0i$; this value is typical of giant planet condensates [1, 7]. In the upper haze the single scattering albedo was set to 0.45. For the lower cloud the single scattering albedo and optical depth were given the wavelength dependant profile shown by Irwin et al. [1], and reproduced in Fig. 3.

For comparison, we also carried out preliminary cloud layer retrievals using Irwin et al.'s bright cloud model variant, in which the single scattering albedo of the upper cloud is 0.85 [1]. The bright cloud model produced a lower altitude upper haze, as well as much greater lower cloud opacities than anything previously reported by Irwin et al. [10]. Together with the propensity of ATMOF to retrieve a thicker lower cloud rather than an upper haze regardless of which of the two models is used, this allowed us to conclude that the standard model was the more appropriate.

The model was calculated at a spectral resolution of 0.01 $cm^{-1}$ before convolving to the instrumental resolution. Upon completion of modelling we checked our model spectra against one calculated with a spectral resolution of 0.001 $cm^{-1}$ and found no discernable difference.

**Preliminary cloud layer fitting**

Our primary goal for this work was to obtain appropriate cloud base pressures and opacities to be used for the high resolution data (with obtaining an initial D/H ratio a secondary objective), to that end we fit these cloud parameters first with the initial values set to those given by Irwin et al. for their bright spot; lower cloud: 2 bar, 0.7; upper haze: $3.5 \times 10^{-2}$ bar, 0.5. As our data has not been flux calibrated we also fit a scaling factor.

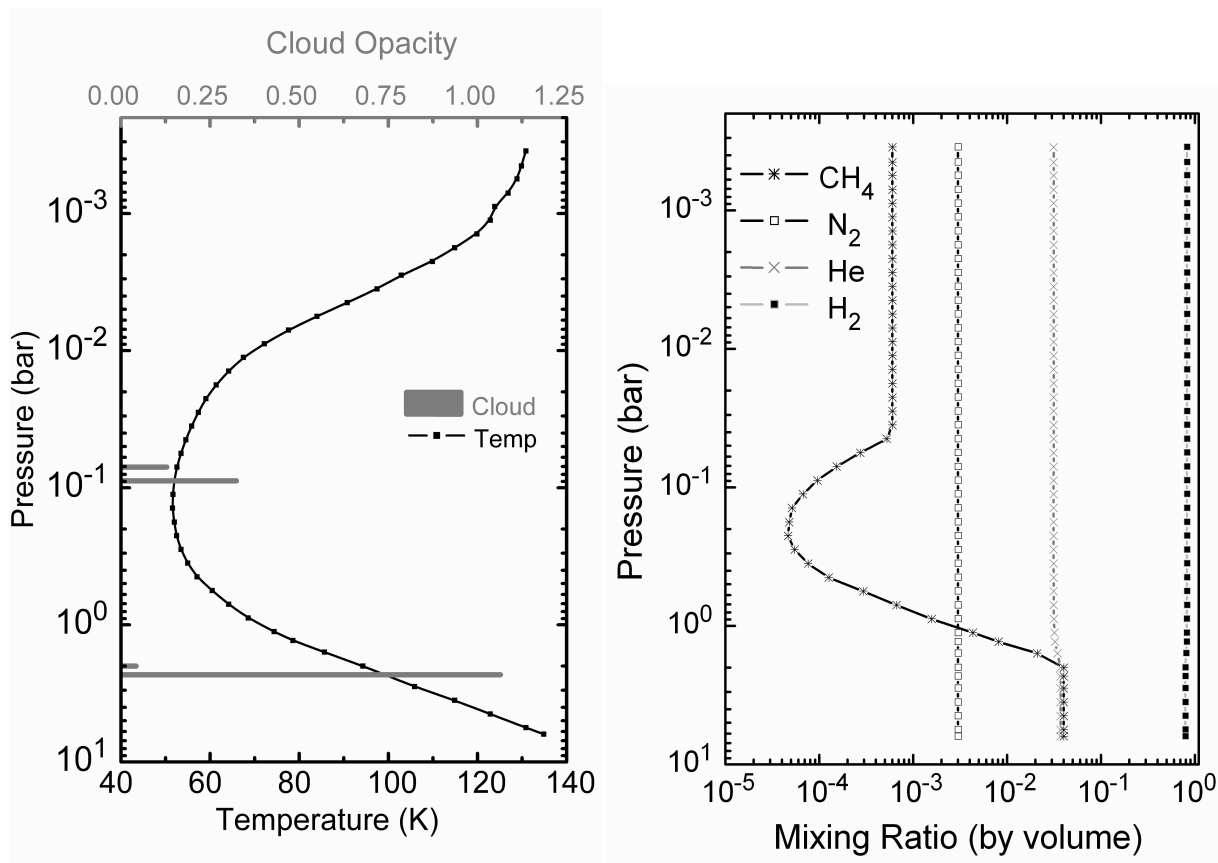

*Fig. 2 The temperature-pressure profile (black) and retrieved cloud heights and optical depths (grey) shown together but represented by separate abscissa (left panel) and the gas mixing ratio profile (right panel) used in the Neptune model.*

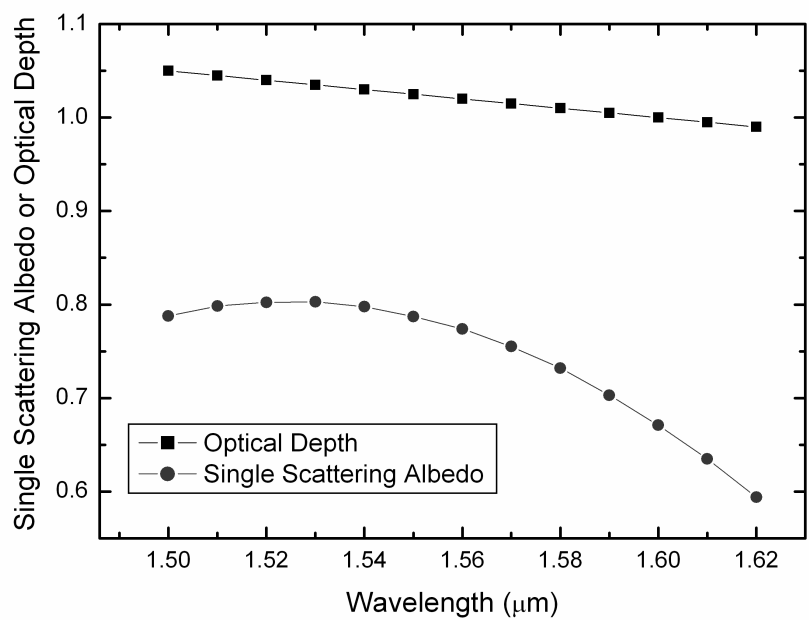

*Fig. 3 Lower cloud properties that vary with wavelength from Irwin et al. [1].*

Our modification to the Irwin et al. cloud model was to split the opacity of both clouds across two of our model layers‡ in a ratio determined based on the nominal pressure of the cloud layer as follows: where the pressure corresponds to the centre of the model layer all the opacity is placed in that model layer; if the pressure is less then some of the opacity is placed in the model layer above, if the pressure is more, some of the opacity is placed in the layer below; the fraction of opacity placed in the model layer above or below is determined by the pressure difference with that of the centre of the layer. This was done in order to supply ATMOF's fitting routine with a continuously varying function, rather than a step-wise one – with which it deals poorly. A base pressure corresponding to the centre of the layer results in all of the opacity being placed in that layer, whereas a lower base pressure resulted in an increasing portion (up to half) of the opacity being placed in the lower layer.

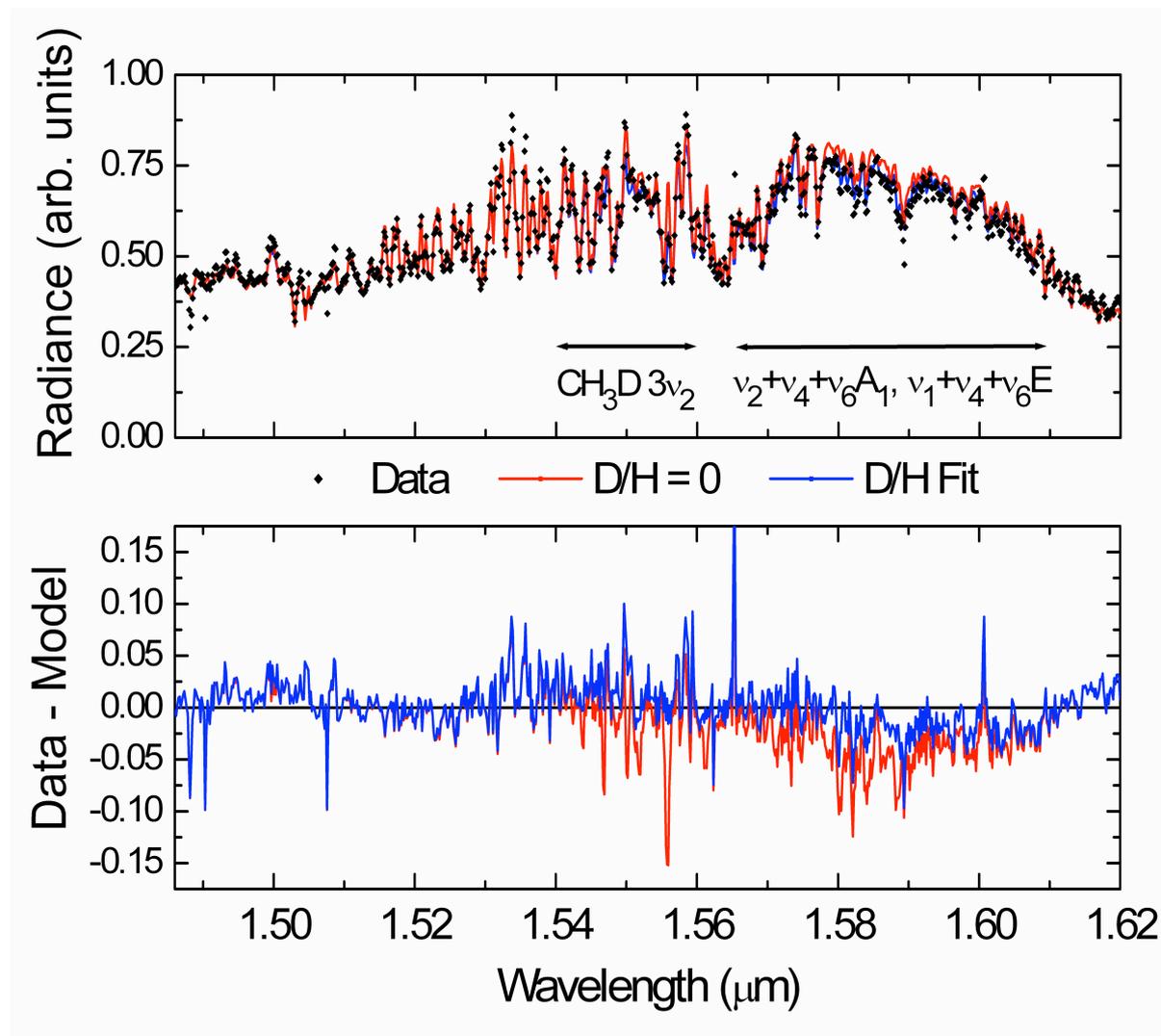

*Fig. 4 Model fit to the data showing the contribution made by $CH_3D$. Marked in the top panel are the regions of absorption due to the $CH_3D$ $\nu_2+\nu_4+\nu_6 A_1$, $\nu_1+\nu_4+\nu_6 E$ and $3\nu_2$ bands.*

**Cloud Layer and $CH_3D/CH_4$ Retrievals**

Only after the initial cloud parameters have been retrieved and the goodness of fit assessed do we fit for the initial D/H ratio as the sole parameter. Then, once ATMOF has achieved a solution, we refit the D/H ratio, along with all the cloud parameters and the scaling factor starting from the previously retrieved values. We find this to be most efficient and less prone to finding an erroneous local minimum as solution.

---

‡ We therefore have two double-layer clouds.

The cloud parameters, of central pressure of the cloud layer and opacity respectively, retrieved for the upper haze were $8.7 \times 10^{-2}$ bar, 0.45; and for the lower cloud 2.0 bar, 1.11, where the break-down of the optical depth into the layers of the model is shown pictorially in the left panel of Fig. 2. The retrieved $CH_3D/CH_4$ was $3.0 \times 10^{-4}$. Fig. 4 shows the fit obtained by the model to our observational data, along with an otherwise identical model devoid of $CH_3D$.

## Discussion and Further Work

Most of the bright features seen by Irwin et al. were the result of thick upper hazes (around 0.1 bar). On the 9th of September, 2011, however, they did observe one feature, like that we report here, that appears to have been due to a thick lower cloud [10]; this cloud was seen at similar latitudes to that which we report here.

This is the first time we have used ATMOF to retrieve parameters for a planet other than Earth. The excellent agreement between our determination of $CH_3D/CH_4$ with that of Irwin et al. [1], and the sensible cloud parameters retrieved by assuming only one unknown parameter – the flux scaling factor, show that ATMOF can be adapted to work well to retrieve basic atmospheric parameters. Further development of this option will include a proper assessment of uncertainties in the derived parameters, that will help in assessing the uniqueness of models under consideration.

As noted by Feuchtgruber et al. [9], recent determinations of the D/H ratio for Neptune (see Table 1), with which our value agrees, correspond to D/H ratios for the proto-planetary ices that are significantly less than those obtained so far for comets, which implies a formation location for Neptune well inside its current orbit [23].

Despite a good agreement between our model and observed spectrum (Fig. 4) the most notable discrepancy is in the failure of the model to match the extremes of depth or height of the finer spectral features, a characteristic shared by Irwin et al.'s fits [1]. To test ATMOF we deliberately limited the number of model parameters we investigated for this work, which limited our ability to precisely fit the data. An obvious candidate for an extra parameter is the deep methane concentration. Here we adopted Irwin et al.'s recommended value of 4%, but the lower value of 2.2% has often been used in the literature based on the work of Baines et al. [1, 18] . Another possibility is an inaccurate line-shape model, as the one we are using was not specifically developed for the conditions of Neptune's atmosphere.

It should also be noted that our model does not contain CO. The value determined in the upper atmosphere, 1 ppm, has previously been used throughout the atmosphere [1]. CO lines if present will show up as distinct P and R branches between 1.570 and 1.582 μm. However, the spectrum is most sensitive to CO around the level of the lower cloud, and Irwin et al. [1] found best agreement in their dark band data (where smaller cloud opacities were retrieved) with no CO at all (but noted that this was affected by the modelled deep methane concentration). Whether this is also the case in a region with a thick reflective lower cloud is something to investigate.

The next step is to flux calibrate our data and determine the uncertainties before re-determining the cloud parameters using the method laid out here, and using these to fit our higher resolution spectra. It is anticipated that doing so will allow us to place tighter constraints on the D/H ratio.

In press, 3/15, to be published 2015. Accepted to the *Proceedings of the 14th Australian Space Research Conference* (14ASRC), Adelaide.

## Acknowledgements

Based on observations obtained at the Gemini Observatory (and processing using the Gemini IRAF package), which is operated by the Association of Universities for Research in Astronomy, Inc., under a cooperative agreement with the NSF on behalf of the Gemini partnership: the National Science Foundation (United States), the National Research Council (Canada), CONICYT (Chile), the Australian Research Council (Australia), Ministério da Ciéncia, Tecnologia e Inovação (Brazil) and Ministerio de Ciencia, Tecnología e Innovación Productiva (Argentina). The observations were obtained under the program GN-2011B-Q-1.